\documentclass[conference]{IEEEtran}
\usepackage{balance}
\usepackage{graphicx}
\usepackage{listings}
\usepackage{xspace}
\usepackage{url}
\usepackage{siunitx}
\usepackage{multirow}
\usepackage{dcolumn}
\usepackage{csquotes}
\usepackage{tcolorbox}

\usepackage{xcolor}
\usepackage{color, colortbl}
\usepackage{adjustbox}
\usepackage{booktabs} % For formal tables
\usepackage{tabularx}
\usepackage{pifont}% http://ctan.org/pkg/pifont
\usepackage{mwe}
\usepackage{graphicx}
\usepackage{subcaption}
\newlength{\thinline}
\newlength{\thickline}
\newcommand{\noind}[0]{\vspace{5 pt} \noindent}
\newcommand{\noindpar}[1]{\noind {\bf #1}}
\newcommand{\sysname}{Privacy Plumber\xspace}
\newcommand{\inspector}{IoT Network Analyzer\xspace}
\newcommand{\etc}{\textit{et al.}\xspace}

\newcommand{\edited}[1]{{#1}}
\ifCLASSINFOpdf
 
\else
  
\fi

\begin{document}
%
% paper title
% can use linebreaks \\ within to get better formatting as desired
\title{Augmented Reality's Potential for Identifying and Mitigating Home Privacy Leaks \vspace{0.4cm}}

% author names and affiliations
% use a multiple column layout for up to three different
% affiliations
\author{\IEEEauthorblockN{Stefany Cruz$^1$, Logan Danek$^1$, Shinan Liu$^2$, Christopher Kraemer$^6$, Zixin Wang$^3$\\Nick Feamster$^2$, Danny Yuxing Huang$^4$, Yaxing Yao$^5$,  Josiah Hester$^6$ \vspace{0.4cm}
}
\IEEEauthorblockA{$^1$Northwestern University, $^2$University of Chicago, $^3$Zhejiang University \\ $^4$New York University, $^5$University of Maryland, Baltimore County, $^6$Georgia Institute of Technology \vspace{0.7cm}}
}

\maketitle

\begin{abstract}
Users face various privacy risks in smart homes, yet there are limited ways for them to learn about the details of such risks, such as the data practices of smart home devices and their data flow. In this paper, we present \sysname, a system that enables a user to inspect and explore the
privacy ``leaks'' in their home using an augmented reality tool.  \sysname
allows the user to learn and understand the volume of data leaving the home and
how that data may affect a user's privacy---\emph{in the same physical context} as the devices in question, because we visualize the privacy leaks with augmented reality. 
\sysname uses ARP spoofing to gather aggregate network traffic information and presents it through an overlay on top of the device in an smartphone app.
The increased transparency aims to help the user make privacy decisions and mend potential privacy leaks, such as instruct \sysname on what devices to block, on what schedule (i.e., turn off Alexa when sleeping), etc.  
Our initial user study with six participants demonstrates participants' increased awareness of privacy leaks in smart devices, which further contributes to their privacy decisions (e.g., which devices to block).

\end{abstract}
%% REDO SECTIONS based on Wifi Privacy Ticker paper
%% http://www.appanalysis.org/jjung/jaeyeon-pub/WiFiPrivacyTicker.pdf
\section{Introduction}
\label{sec:intro}

%Internet-connected devices are projected to reach over 75 billion worldwide by
%2025, punctuated by a proliferation of devices in the ``smart home''. Though
%the main purpose of Internet connected devices is to make our daily lives
%simpler, connected smart homes present new privacy risks, as information can
%be leaked without explicit consent of the user.  Third parties can gather
%deeply invasive information about a user based on their interactions with
%smart devices in a space. Data collection is common (and often legal) and
%often includes reputable device software vendors, device manufacturers, and
%third-party app developers~\cite{acar2018web}. 

The increasing adoption of Internet-connected smart devices has brought huge improvements to our lives. Yet, these devices also raise significant privacy concerns from their users, such as sensitive data collection~\cite{zheng2018user, zeng2017end}, data sharing~\cite{zeng2017end}, and data misuse~\cite{malkin2018can, malkin2019privacy, mohcharacterizing}. Literature has suggested many types of privacy risks associated with smart devices. For example, some seemingly innocent data, such as the network traffic shapes and patterns of smart devices, may reveal sensitive personal information, such as users' daily schedule, their gender, date of birth, social security number, location, and behaviors~\cite{apthorpe2017smart, alrawi2019sok}. 
% This personal
% information is then collected and often used for monetization and product
% targeting purposes by the companies who distribute these Internet-connected
% devices, the application developers, and other third parties~\cite{acar2018web}.  

However, many risks are not obvious to users due to the opaque nature of the data practices of smart devices; the average users lack an understanding of how their data is collected, processed, and shared~\cite{zeng2017end, yao2019defending,lipford2022privacy}. Prior research has proposed various ways to increase users' awareness of the data practices in smart homes, such as data dashboards, mobile phone apps, ambient light and sounds, and so on~\cite{thakkar2022would,huang2020iot, castelli2017happened, jin2022exploring}. Some other mechanisms (e.g., IoT Inspector~\cite{huang2020iot}) focus on specific aspects of the data practices and present network traffic data to users so that they can access first-hand data of the data flow in/out of smart devices. Yet, most mechanisms we know decouple such transparency from the device themselves---i.e., users need to learn about the data practices separately from the smart devices---making the information less intuitive to consume, especially for the average user. In addition, these mechanisms do not provide users with the ability to take action if they notice unexpected data practices (e.g., blocking the data from being sent out to third parties).

\begin{figure}[t] %tbp \centering
    \includegraphics[width=0.5\textwidth]{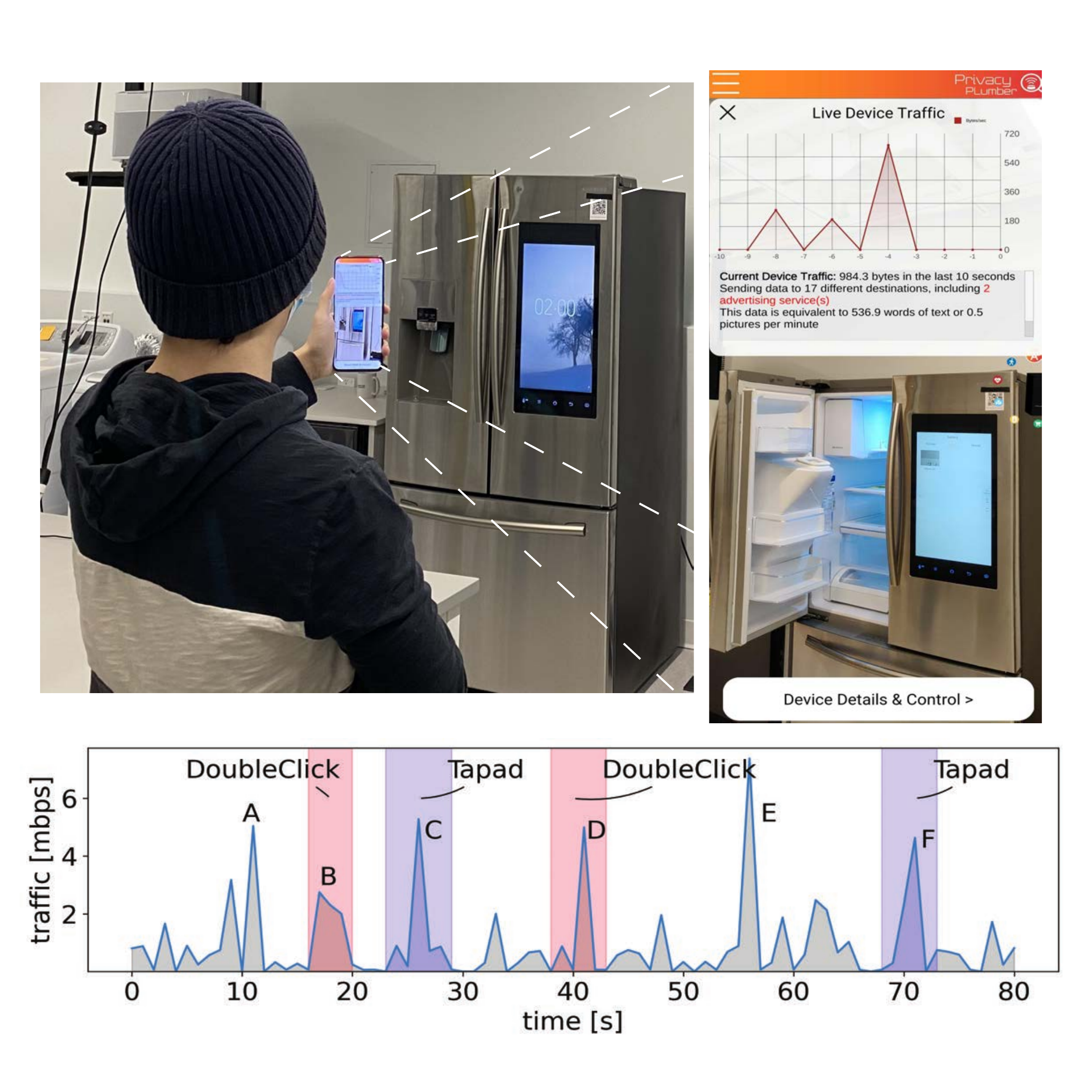}%
    
    \caption{Privacy Plumber lets a user find and mitigate potential privacy violations in the
    smart home. The figure shows a user walking around the smart home
    and inspecting the traffic and trackers coming out of a Samsung Smart Fridge
    using the Augmented Reality enabled app.
    Furthermore, (not shown in the picture above) users can use built-in, infrastructure-free controls to limit traffic of
    devices to times of day---without requiring any additional hardware or modifications to the network. The graph shows the actual network traffic as the user interacted with the Smart Fridge: A: turning on the ice maker; B: browsing recipe; C: browsing
    goods; D: interacting with the Bixby voice assistant of the fridge; E: opening the fridge door; F: adding items to the shopping list.
    During these interactions, the Smart Fridge communicated with various advertising and tracking services, such as DoubleClick and Tapad.} \label{fig:overview}
    
\end{figure}

In this paper, we focus on the data flow in and out of smart devices. We build a proof-of-concept smartphone-based augmented reality system called \textbf{Privacy Plumber} to increase users' awareness of the data flows of smart devices and provide them with controls to block certain data flow if needed. We focus on data flow rather than other aspects of data practices (e.g., types of data being collected) mostly due to practicality and feasibility reason, as we can reasonably capture data flow and identify its source and destination using ARP spoofing~\cite{huang2020iot}. In addition, from the smart devices' perspective, these devices have multiple tiers of software, all of which entail some type of tracking. Such tracking is generally embodied in the data flow. We use augmented reality to visualize data flows \emph{in the same physical environment} as the devices in question; this method could potentially help users establish a connection between the devices and their data flows in the same context. Users' proper understanding of data flow may help them understand the privacy implications of devices such as smart TVs~\cite{mohajeri2019watching}, voice assistants~\cite{huang2020iot},
children's toys~\cite{chu2018security}, security cameras~\cite{mare2020smart, pierce2020sensor}, and smart light bulbs~\cite{bombik2022multi}.

% The device manufacturer gathers usage information, the applications hosted on
% the device (e.g., Music Player, Recipes) collect location, user actions, apps used, and so forth. 
% These privacy leaks occur in
% other smart home devices, including smart TVs~\cite{mohajeri2019watching}, voice assistants~\cite{huang2020iot},
% children's toys~\cite{chu2018security}, security cameras, and smart lights.

The development of Privacy Plumber is inspired by the following three gaps in the literature. First, the data flows of smart devices are opaque and not visible to users. Second, existing tools to monitor network traffic of smart devices (e.g., IoT Inspector~\cite{huang2020iot}, open.Dash~\cite{castelli2017happened}) require a certain level of technical knowledge to be able to interpret the results---not to mention that the results are often decoupled from the physical environment where the smart devices are situated. Oftentimes, the results are presented on, for instance, dashboards on computers or phones, where there is a disconnection between the visualization of data flows and the smart devices that create the data flow. Third, existing tools or mechanisms do not provide users with the ability to control unnecessary or unexpected data flows. With Privacy Plumber, we aim to bridge the gaps and increase users' awareness and control of the data flow in smart devices. 
% \danny{please read this paragraph; i just expandeded it}

% \yaxing{Maybe add a little bit more of the system} \danny{The above seems enough to me. What else do you want to add?}

Privacy Plumber uses augmented reality (AR) techniques and visualizes real-time network traffic flowing in and out of smart devices through an overlay. It allows users to find potential privacy leaks in their homes by pointing the AR-based app at smart devices. As shown in Figure~\ref{fig:overview}, the app  adds an overlay on top of the smart devices in which it displays a real-time data flow based on the network traffic with the necessary information for users to understand it. We chose to use AR because, as privacy is highly contextual~\cite{nissenbaum2009privacy}, it can provide strong contextual connections between the actual real-time privacy leaks, and the user actions (or inaction). This allows the smartphone to function as a viewfinder into the invisible world of data flow and identify potential privacy violations. 
% It also gives a heads-up display of information visuals on potential privacy leaks and controls. 
The smartphone application relies on a companion software tool hosted on a laptop or desktop on the same home network. This tool discovers smart devices in a user's home, intercepts their traffic via ARP spoofing~\cite{whalen2001introduction}, and analyzes the data flow (e.g., what traffic is leaving the home over time)---without requiring the user to modify their network settings or install additional hardware. When users would like to take action and block certain data flow, ARP-spoofing is used again to jam specific devices' traffic (thereby blocking the device) at the time of day set by the user.

We build a proof-of-concept prototype and conducted a pilot study with 6 participants in our lab to collect their feedback on the prototype. Our initial findings have suggested that \sysname helped participants  understand the network traffic, increased their awareness of potential privacy violations, and helped them make more informed decisions on how to handle IoT devices.

% It allows users to find potential privacy leaks in their homes by pointing the AR-based app at smart devices. The app will add an overlay on top of the smart devices in which it displays a real-time network traffic trend with the necessary information for users to understand it. We build a proof-of-concept prototype and conducted a pilot study with 6 participants in our lab to collect their feedback on the prototype.  

% Because smart devices are everywhere, and the context and environment of the device matter in the leaks it can create, we implement the user-facing portion of the tool on a smartphone. 

This paper makes three contributions. First, to the best of our knowledge, Privacy Plumber is the first mechanism that provides users with real-time information on the data flow of their smart devices. This paper proves the possibility of using AR-based technology as a viable option to increase users' awareness of the data flows of smart devices. Second, our initial evaluation shows promising results, indicating users' potential acceptance of these technologies. Third, we summarized lessons learned from the pilot user study to inform the design and development of future systems that aim to improve users' awareness of data practices in smart homes.

\section{Background and Related Work}
In this section we discuss related work seeking to understand or discover privacy leaks, and the tools that exist to help users understand and mitigate them. %Much of the related work is complementary or enhances the usefulness of \sysname. 
%For example novel machine learning based approaches to extract activities from network traffic and identify devices, or exploring design spaces of privacy tools. 
\sysname is meant to to provide a handheld and zero-cost inspection and experimentation tool for privacy leaks of nearby smart devices in the home, and a straightforward and low burden method for mitigating those leaks. 

\begin{figure}[!t] %tbp
\centering
\includegraphics[width=0.5\textwidth]{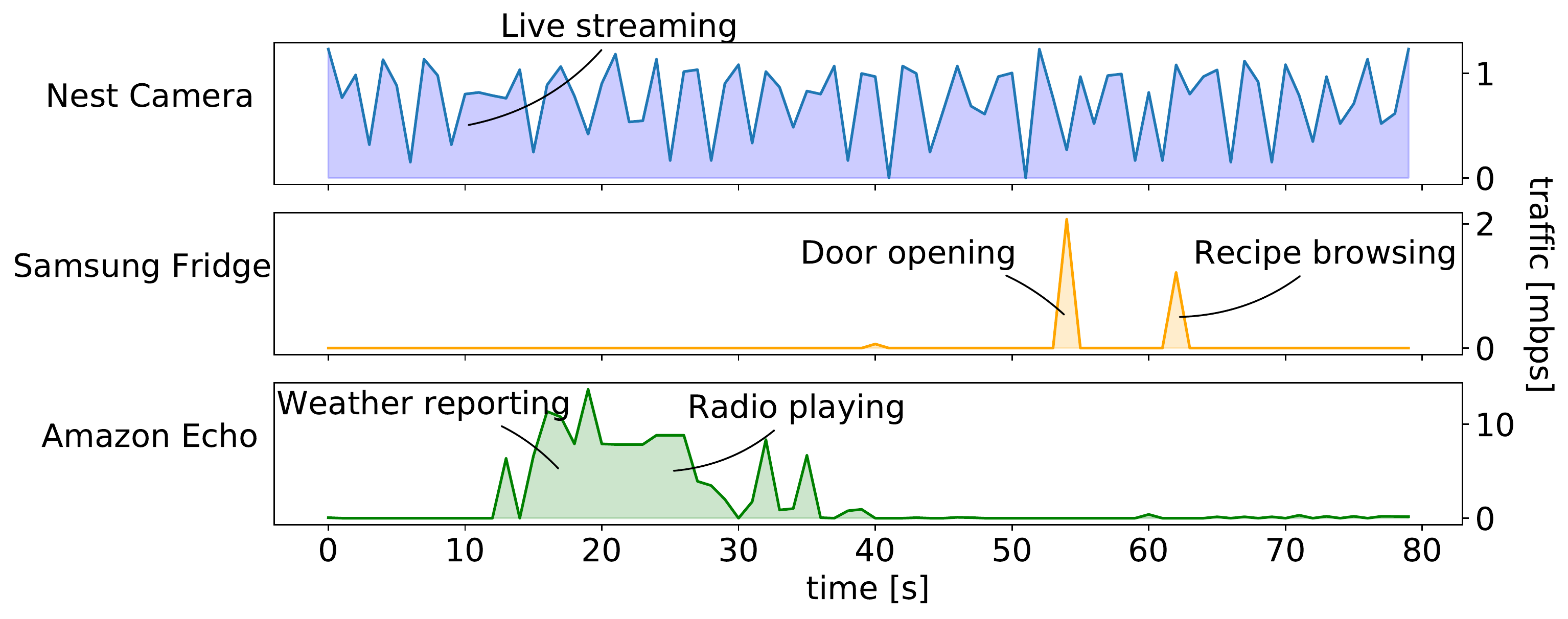}%
\caption{Outbound network traffic from various smart home IoT devices: a Nest Camera, an Amazon Echo, and a Samsung Smart Fridge. Traffic increases or provides a fingerprint for many types of seemingly benign actions, creating a privacy leak. Current systems do not provide real-time context or ability to experiment with these devices, nor control their leakage.}
\label{fig:network-traffic}
\end{figure}

%\noindpar{Network Traffic in the Smart Home.} Smart devices give off digital exhaust which can be used by 3rd parties including a user's Internet Service Provider, advertisers, device manufacturers, and others, to fingerprint activities and get sensitive information. Shown in  Figure~\ref{fig:network-traffic} is the network traffic and trackers of various smart home devices.This network traffic forms the basis of most leaks.
\subsection{Privacy Issues in Smart Home}
Over the decades, privacy issues have been deeply disclosed in smart home, such as transparency of data collection, data sharing, and accessibility~\cite{lin2016iot,yao2019defending, worthy2016trust, jin2022exploring, zheng2018user, naeini2017privacy,yao2019defending}. Some smart home devices have always-on sensors that capture users' offline activities in their homes and transmit relevant information outside of the home, especially for cloud services run by device manufacturers~\cite{apthorpe2017spying}.

In the meantime, users are concerned about leaks of sensitive information~\cite{malkin2019privacy, zeng2017end, mcreynolds2017toys}, such as visual and auditory information which they see as private~\cite{malkin2019privacy, mcreynolds2017toys}. Thus, users have a strong desire to protect themselves against such recordings being accessed without their permission~\cite{naeini2017privacy, lee2016understanding}. However, some information users perceived as not very sensitive also lead to privacy leaks. For example, the home temperature could be used to determine whether a house is occupied or not, as a precursor to burglary~\cite{lin2016iot}.

In fact, smart devices give off digital exhaust which can be used by third parties including a user's Internet Service Provider, advertisers, device manufacturers, and others, to fingerprint activities and get sensitive information. Shown in  Figure~\ref{fig:network-traffic} is the network traffic and trackers of various smart home devices.This network traffic forms the basis of most leaks.

\subsection{Tools for Enhancing Smart Home Privacy}
Most related to \sysname are tools that watch or monitor network traffic in the home and provide something of use to the user, whether visualization and information, education, or a mechanism for control.

Sophisticated, technically literate users can use systems that block
advertising and tracking domains (e.g., PiHole~\cite{RaspberryPiPoorNetworkingPerformance2} and
pfSense~\cite{pfsenseArticle}), but these methods are bespoke and often require
additional or dedicated hardware (e.g., Raspberry Pi for PiHole, and a
supporting custom router for pfSense).  Other tools have provided insight into
what might be exposed from web-browsing activities, including WiFi privacy
Ticker~\cite{consolvo2010wi}, but do not consider or scale to the new problems
of connected devices with physical sensors and abilities in a space.
Aretha~\cite{seymour2020informing} explores this tool space and proposed (but
did not deploy) a simple firewall-based control mechanism.  Aretha presents
data in aggregate instead of contextually and in real-time.  None of these
techniques investigate a range of IoT devices, usually constrained by studies
with participants in their own homes, in a time when smart home adoption is
low (Aretha had three participants, and only one had more than a phone,
tablet, and Alexa).  None of these techniques develop a scalable (no
additional hardware required) way to interpret privacy leaks and control them.
Emerging smart devices are highly contextual and location sensitive, an Alexa
in the bedroom versus the kitchen  has different privacy exposure (i.e. the
former gives sleep times, the latter exposes eating habits).  Moreover,
tracking these devices' privacy exposures presents a technical challenge because
the traffic is not centralized through a web browser or laptop.  A tool is needed to visualize privacy leaks from smart devices in real-time and in context, educate users on the consequences of these leaks, and
provide control mechanisms for partially mitigating these leaks.

Wifi Privacy Ticker~\cite{consolvo2010wi} demonstrated a first method for improving the awareness of users in terms of privacy by providing a count of the amount of sensitive data that was being transmitted unencrypted over the network awareness. By seeing this in real-time, users could adjust their behavior or find encrypted means to browse the web. 
Of course, this ticker was developed well before the current generation of smart devices, however the underlying concept of surfacing the invisible privacy leaks remains the same for \sysname, but for smart devices. 
Xray-refine~\cite{van2018x,van2017better} provided smart phone users a means to visualize their exposure profile, based on the duration of app use.
This method was an educational solution, but users had to adjust behavior to work around the constraints of the apps they were using, in some cases, opting out of apps to reduce privacy exposure.

Finally, recent work like Aretha~\cite{seymour2020informing}, \edited{PriView~\cite{priview}, Lumos~\cite{lumos}}, and IoT Inspector~\cite{huang2020iot} look at making usable visualizations and mechanisms to understand and interpret data coming from smart devices in the home.
IoT Inspector is a simple-to-install desktop application that uses ARP spoofing to gather network traffic on the Wifi network of the desktop/laptop. 
This information is sent and collated at a server, and then viewed online by the user, listing different trackers and websites that are attached to smart device usage. Because of the ease of installation and no extra hardware requirement, IoT Inspector was deployed by thousands of users.

In comparison, Aretha is a part research tool, part exploratory users tool for exploring a design space of privacy tools and controls. 
Aretha helps users become aware of the network traffic flows in their homes while also educating users to regain their privacy in the connected home.
Aretha suggests the use of firewall mechanisms controllable by the user, but does not implement them.
Aretha, owing to a hardware requirement (a device must be attached to the Wifi router in the home) was only deployed in three homes, compared to the massive scale deployment of IoT Inspector. 
\edited{Similarly, PriView also has a hardware requirement; its users need to have dedicated external thermal cameras (e.g., FLIR One~\cite{priview}) attached to their phones. For Lumos, there is no special hardware environment, although the focus is more on identifying hidden smart devices rather than analyzing the network traffic for privacy leaks.}

\sysname is not meant as a research tool or a design space exploration tool. It is meant as an actual, real world system with a focus on scalability and ease of deployment in any home, similar to IoT inspector.
Unlike both IoT inspector and Aretha, \sysname provides \textit{real-time and contextual visualizations of privacy leaks}, real-time ability to plug those leaks (as well as automated rule setting for plugging leaks), and enables experimentation in real-time.

Finally, other significant measurement campaigns on in-home traffic have been conducted, focusing on the Wifi network itself or devices in the home~\cite{razaghpanah2018apps, kreibich2010netalyzr}. These have usually been for research purposes and need finding and are useful for informing the design of \sysname, but are not necessarily tools for controlling smart home device privacy.

%\josiah{Add other works liek this, home snitch etc}

\subsection{Determining Home Activities from Network Traffic}
Complementary to \sysname are other works which demonstrate the ability to infer activities from network traffic: whether on a phone, smart device, or laptop~\cite{apthorpe2019keeping}.
By analyzing the patterns of network traffic in the home, occupancy, habits such as sleeping, watching TV, listening to music, and sometimes preferences, can all be determined. 
\textit{HomeSnitch}~\cite{oconnor2019homesnitch}, \textit{Peek-a-boo}~\cite{acar2020peek}, and \textit{HoMonit}~\cite{zhang2018homonit} all utilize machine learning with varying degrees of success to identify activities in the home from network traffic.
Other tools utilized for monitoring Internet connected smart devices in the home, IoT Sentinel~\cite{miettinen2017iot} and
IoT Sense~\cite{bezawada2018iotsense}, have shown that particular devices can be fingerprinted by their traffic patterns. Enabling another way for an ISP or third party to determine the activity in the home.
Each of these systems and methods are complementary with \sysname; inferred activities from traffic would be useful to surface in \sysname for the user to understand privacy exposure and know when to mitigate it, and device fingerprinting provides a way for zero-registration or setup of \sysname in a home.

\subsection{Challenges: Contextual, Real-time Privacy Understanding and Control in the Home}\label{ss:challenges}

Despite the diverse work in the smart home privacy space, significant gaps and challenges remain, which we detail below.

%Studies have shown how the design of IoT devices affect not only its owners but people within its vicinity, "bystanders"[cite tangible]. In Tangible Privacy: Towards User-Centric Sensor Designs for Bystander Privacy, Ahmad et al advise that the design of sensors embedded in IoT devices should be more "intuitive and tangible" to "convey a clear and definite sense of privacy to bystanders so they can be assured of their privacy". While this is a noble goal (tangible privacy) for future IoT devices, it does not help the current slew of devices already in the wild. Some other mechanism is needed to make privacy real for these devices, and then go further, giving back control of privacy to the user.

% Integrate
% Contextual information, interaction, needs to be mobile and real-time
% Need controls, but also need to know how to use it, and see its effectiveness

%\noindpar{C1: Devices dont tell you what they are doing.}
%1. 4.1 Uncertainty about the device states 

%\danny{That's what IoT Inspector does, although its presentation could be too technical for users and they didn't do any user studies on how effectively they can communicate the security and privacy risks to the users.}

\noindpar{C1: Users can't model what devices are doing, especially without context.}
With tools like IoT Inspector, a user might be able to count the number of trackers and advertisers contacted in a day from the sum of their interactions with smart devices.
But how can a user know that turning on the NPR podcast on their smart fridge will send thousands of bytes of information to Bloomberg News for advertising purposes? How can they know that turning on the device sends a short burst of traffic?
Users know that data captured will often be used for advertising, which often generates an adverse reaction~\cite{ur2012smart}.
However, with smart devices, it is not always clear what actions or contexts trigger data being transmitted~\cite{feng2021design}.
Things like Privacy labels for websites and smart devices are meant to give a method for scoring devices privacy~\cite{shen2019iot,kelley2009nutrition}.
However, these are static representations of the privacy exposure of a device.
With tools like IoT inspector and Aretha, \textit{aggregate} views of data are seen (as opposed to \textit{real-time} views), not associated with very fine user actions: like the turn on the light, say command to Alexa, or open the fridge door.
Because of this granularity, the mental models of what devices are doing, and what they are sharing, are very perplexing.
Privacy tools must address this lack of action mapping to network traffic, enabling contextual integrity~\cite{nissenbaum2009privacy} in real-time.

%\danny{If we can argue that AR can improve users' mental model of privacy, that'd be gold! This is because traditional techniques, including IoT Inspector, could be two dimensional and abstract, whereas an AR approach integrates privacy leakage into a real-life three dimensional space blah blah blah. We need to make it super crisp that AR helps improve users' mental model (maybe more so than other conventional methods)... I don't know the relevant literature in AR though... Most work in the space falls into two categories: (i) providing otherwise hidden information, such as IoT Inspector and the Lumen paper (privacy for mobile apps, NDSS 18); (ii) polishing the information for general consumers (such as Lorrie Cranor's work on privacy labels. Our AR approach not only offers (i) and improves (ii). In other words, whereas Lorrie Cranor offers nice looking infographics on privacy on a two dimensional space, we bring the information to a 3-D space (actually 4D if you consider time). If we make this argument, our test case would have to include 2D infographics (to mirror what Lorrie did), as well as our 3D/4D solution.}

\noindpar{C2: Users don't have intuitive methods to control the privacy ``valve''.}
Users want devices that provide helpful features, but they do not know the cost of this ease. One option is to just unplug the device; however, this is all or nothing. Users need a way to valve the privacy flow to something they are comfortable with, or to at least be able to analyze the tradeoffs~\cite{tabassum2019don}.
Making privacy more "tangible"~\cite{ahmad2020tangible} is one way this can be done; where the privacy leaks are more visceral.
% Not sure how to say this, but firewalls are not enough
Selective firewalls (such as pfSense~\cite{pfsenseArticle}), or other more fine grained network mechanisms may provide a means to control the privacy valve, but this must be intuitive and understandable to the user, and they must be able to actually "see" the effect of turning this valve.

\noindpar{C3: Smart devices are context (location, time, action) dependent.}
Smart devices are necessarily scattered around the home; and this will continue as more devices become intelligent, and more applications are explored. Watching a desktop or laptop traffic meter and figuring out which device in which room is doing what at which times, is mentally trying for the user and disassociates the device from the physical space that defines its context and use.
Just like when trying to find leaks in pipes, physical proximity is required. Handheld inspection tools provide mobility, and enable in-situ fixing and experimentation.

\noindpar{C4: Users can't experiment.} Indeed, because of contextual changes in how private information is leaked, experimentation is difficult with existing tools that generally provide traffic summaries. Interactions with smart devices can last only a few seconds. Enabling a user to experiment with different actions and uses of a smart device, and then see the associated network traffic in real-time, would provide a powerful way to build a mental model. However, providing an ability to experiment is challenging with the current suite of tools.

\noindpar{C5: Technical challenge of scalability and deployment.}
If a privacy tool is to be useful and translate to the general public, it must be hardware free, or at least trivially easy to deploy to enable scalability and broad adoption.
Commercial products like \url{fing.com} embed all functionality in a single phone application. Large scale deployments like with IoT Inspector are enabled through a desktop application that is easy to install.
However, these methods do not provide controls since that is technically difficult to do without custom hardware put between the Wifi endpoint and the user. On the other hand, hardware requirements or custom install procedures reduce the deployment size of tools like Aretha, or narrow the user base by requiring technical ability, as with PiHole.
It is not clear how to implement mechanisms of control without changing the Wifi network and infrastructure. To create scalable, user-centered, novice friendly privacy tools, mechanisms for enabling control of smart device traffic without hardware intervention must be developed.

%\noindpar{Summary of Challenge: Users do not understand the privacy / utility tradeoffs.}
%\josiah{Summarize here basically, what this means, and how we can build this end to end system to address. Then make a call to arms: we need a NEW system to address this!}

\begin{figure*}[htp]
    \centering
    \includegraphics[width=0.7\textwidth]{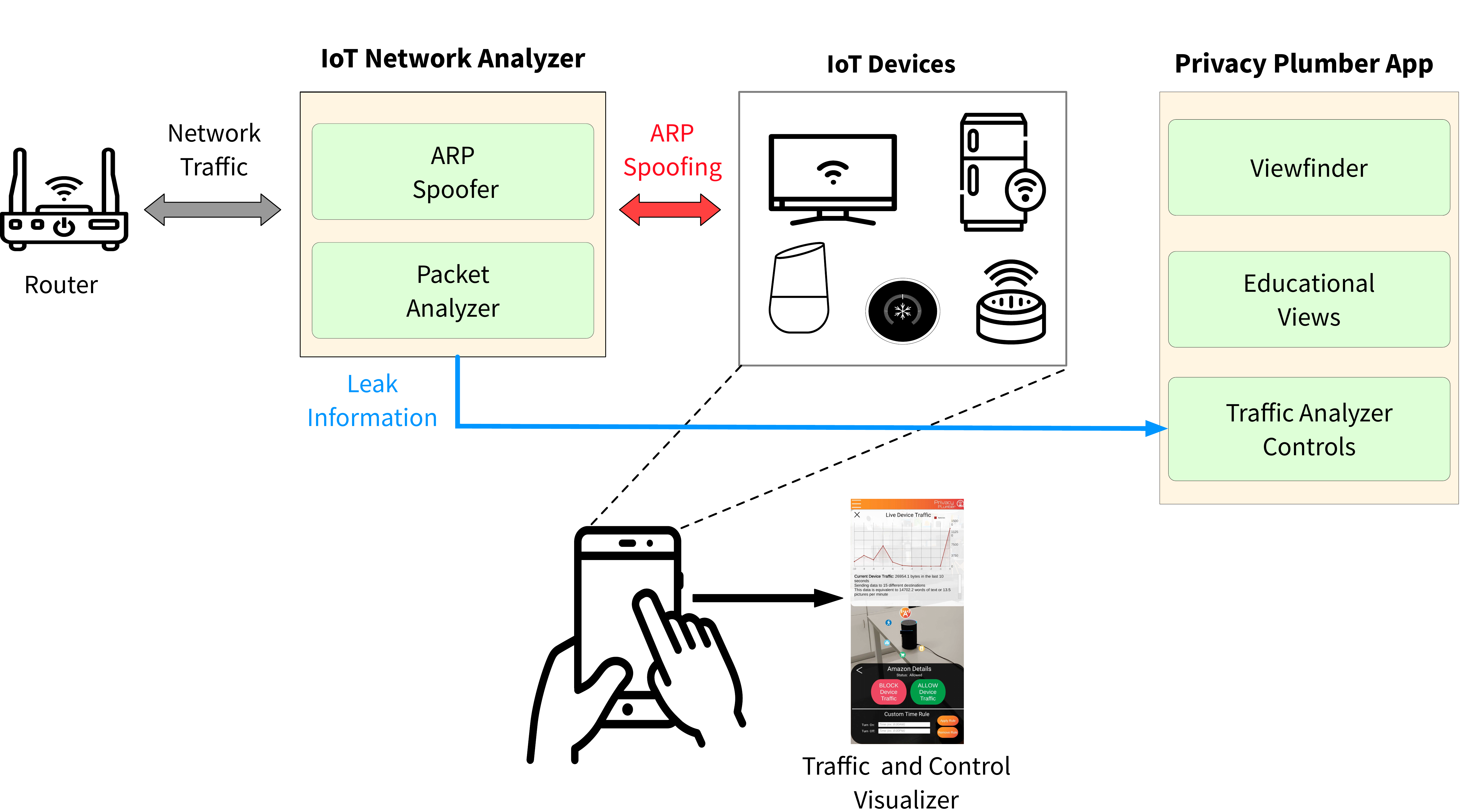}
    \caption{ System diagram of \sysname including the IoT Network Analyzer and \sysname mobile application. IoT Network Analyzer runs on a computer that is connected to a user's router. IoT Network Analyzer automatically discovers and captures IoT devices on the same network using ARP spoofing. Privacy Plumber connects with IoT Network Analyzer to present the network analysis in AR. The user can then examine their devices' network traffic and control when they want their devices to be on or off.}
    \label{fig:sys-overview}
    
\end{figure*}

\section{System Design}

We present \sysname as a proof-of-concept and end-to-end system to address the challenges listed of scalable and general population serving privacy tools for emerging smart homes.
\sysname is inspired by various handheld tools for identifying and fixing faults in large and complex systems.
For example, acoustic leak finding has been used for decades to localize leaks in gas and water pipelines.
Handheld oscilloscopes, multimeters, and RF Spectrum Analyzers have helped engineers debug problems in large electrical systems.
These handheld devices make the invisible signals visible and interactive. They allow real-time experimentation and debugging.
Inspired by these devices, \sysname is designed to offer a general user a level of insight and control of the invisible privacy leaks that are rampant in Internet-connected smart devices in the home.
\sysname is composed of two pieces as shown in Figure~\ref{fig:sys-overview}:

(1) the \textbf{\inspector}, a desktop application which collects real-time data on smart devices on the shared Wifi network, and provides an infrastructure and hardware free mechanism to block arbitrary devices traffic, and;

(2) the \textbf{\sysname} phone application, which serves as a viewfinder or inspector for any smart devices in view, and presents data from the desktop application, including device 
 network traffic and potential privacy leaks to the users, along with educational content matched to what is known about the device, all in real time.

\noindpar{Overview of Usage.} 
A user would first download, install, and run \inspector on their computer and the \sysname app on their mobile phone, such that both the computer and the phone are on the same local area network. Let us assume that the user is interested in inspecting a smart device like an Amazon Echo. While running the \sysname app, the user points the phone camera to Echo and speaks a voice command (e.g., ``Alexa, what is the weather?'') \inspector captures all network traffic between Echo and the Internet, analyzes the packets, and identifies destinations that are third-party advertising and tracking companies. The \sysname app extracts this information from \inspector and visualizes key statistics for the user---such as real-time bandwidth usage of the device and the number of advertising and tracking services contacted---as an overlay in the AR view.
    
\begin{figure*}[!htp]
    \begin{subfigure}[t]{0.2\textwidth} 
     \includegraphics[width=\textwidth, height=3in]{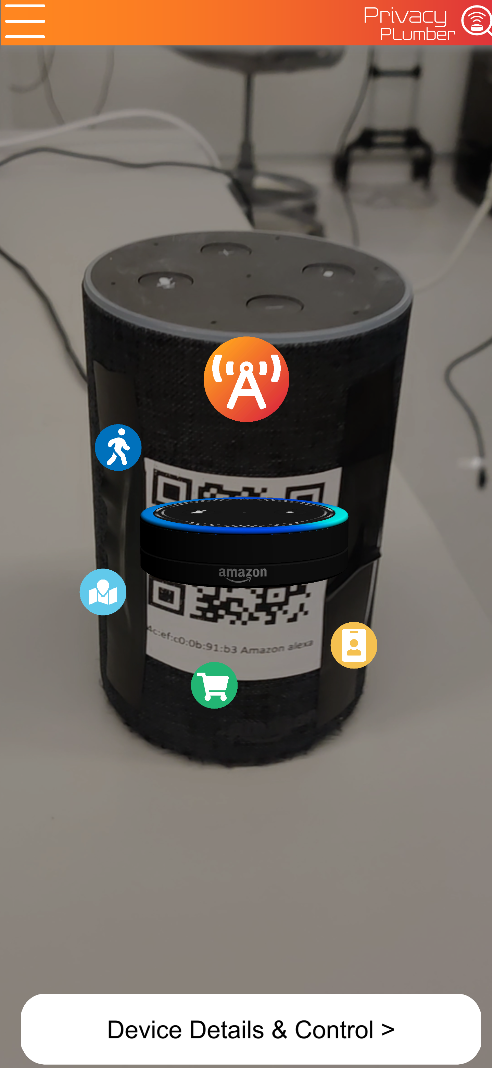}
     \caption{View finder}
     \label{fig:scan}
   \end{subfigure}
   \hfill
   \begin{subfigure}[t]{0.2\textwidth} 
     \includegraphics[width=\textwidth, height=3in]{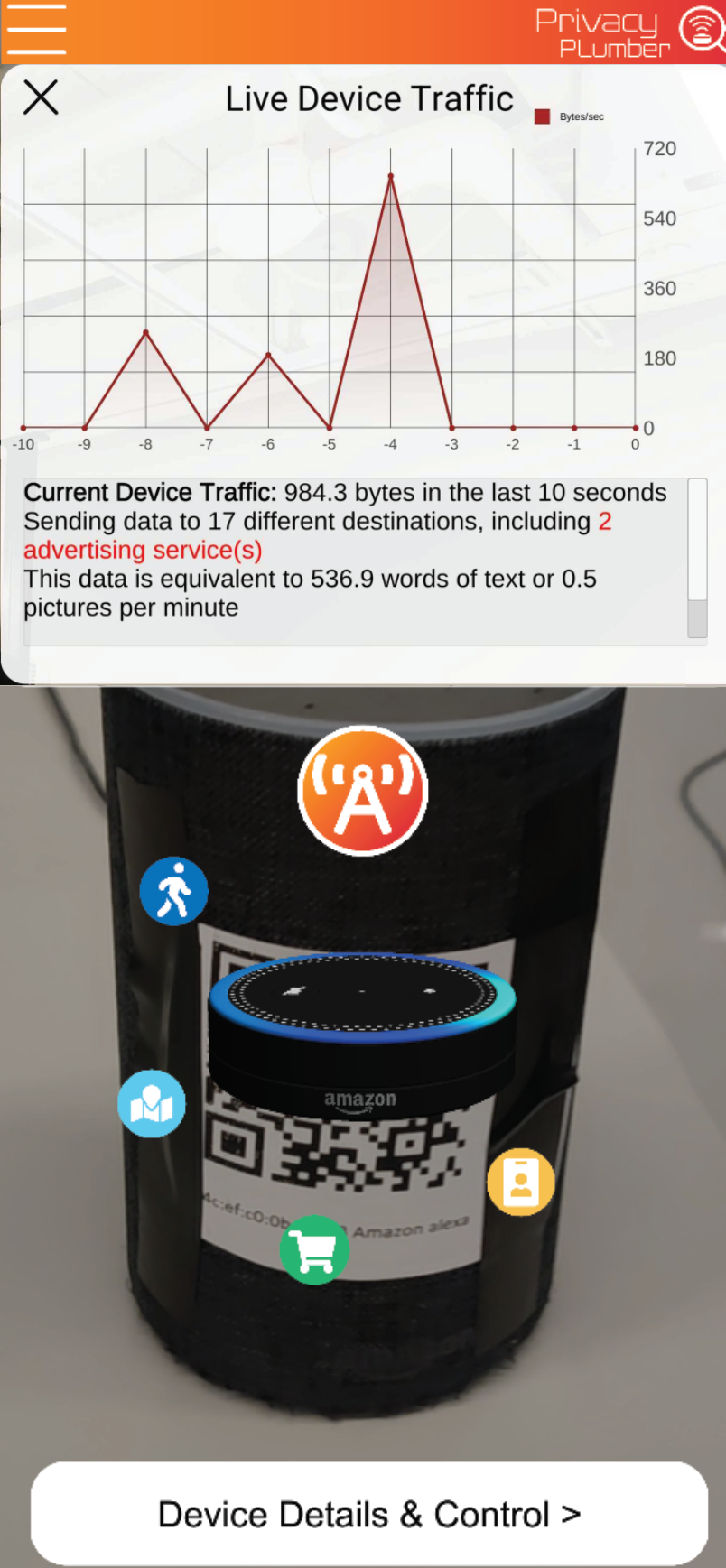}
     \caption{Traffic view}
     \label{fig:live}
   \end{subfigure}
   \hfill
   \begin{subfigure}[t]{0.2\textwidth}
      \includegraphics[width=\textwidth, height=3in]{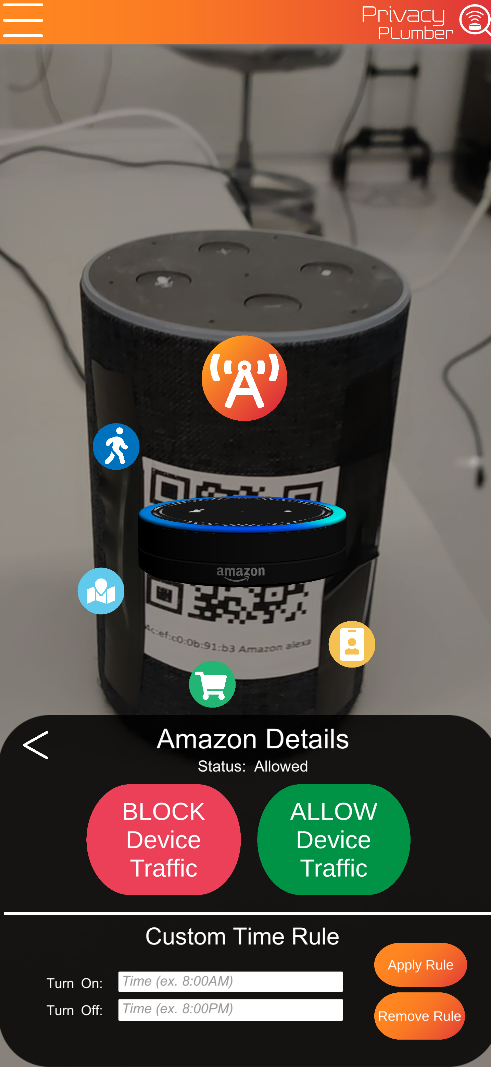}
     \caption{Controls}
     \label{fig:control}
   \end{subfigure}% 
   \hfill
   \begin{subfigure}[t]{0.2\textwidth}
     \includegraphics[width=\textwidth, height=3in]{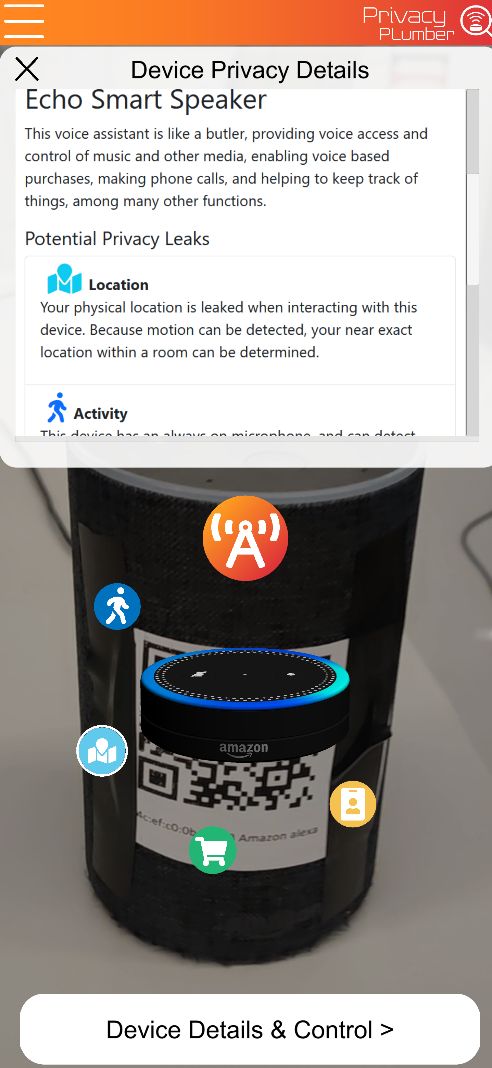}
     \caption{Education}
     \label{fig:education}
   \end{subfigure}
   \caption{ Illustration of mobile application design. (a) Device recognition with interactive menu. (b) Live traffic inspection. (c) Rule-based device traffic control (i.e., blocking and unblocking). (d) Educational material on privacy details.}
   \label{fig:Designs}
\end{figure*}

When the user points the phone camera at a device, the \sysname app does not recognize devices with computer vision algorithms. Instead, for the purpose of this prototype, we print a QR code on each IoT device. The QR code includes the device's MAC address, its name, and the manufacturer. The app uses the phone's camera to scan for the QR code, identifies the device based on the QR code, and displays the device with a dial menu around it (see Figure~\ref{fig:scan}). The options in the menu allow the user to see the outbound traffic from the device as well as read a brief article stating what types of information the device may be tracking. The user may also use the Device Control menu (Figure \ref{fig:control}) to manually block or allow traffic from the device. Future versions of the app will use computer vision to recognize devices; see the discussion in Section~\ref{sec:discussion}.

%\josiah{HANDHELD and PORTABLE. Scalable and REAL WORLD applicable. This could be deployed to anyone with a phone and a laptop! Unlike Aretha and others.}
%Or signal vector analyzer
%Or handheld oscillosope?

\noindpar{Privacy Threat Model and Assumptions.} 
We assume that a user's privacy may be \textit{potentially} violated if an IoT device exhibits either or both of the following behaviors. In \textbf{Threat 1}, an IoT device could contact an advertising and tracking service on the Internet. In \textbf{Threat 2}, an IoT device could be sending out network traffic to hosts on the Internet when the user does not expect any network activities---for example, when the user is not interacting with the device. 
    
We design both the \sysname app and \inspector with this privacy threat model in mind. \inspector captures packets, analyzes the headers,  identifies the destination hosts (based on the IP addresses, domain names, and the TLS Server Name Indication fields), and determines if a destination host is an advertising and tracking company. The \sysname app displays the number of advertising and tracking services (e.g., the red text below the graph in Figure \ref{fig:live}), thereby helping users toward identifying \textbf{Threat 1}. Based on the byte counters from \inspector, the \sysname app also shows a bandwidth graph that plots the bytes sent per second over time (e.g., the time-series graph in Figure~\ref{fig:live}).  This graph could help users correlate network activities with human interactions---or the lack thereof---with given IoT devices and thus identify possible instances of \textbf{Threat 2}. Note that \inspector does not parse the payload of packets to discover sensitive information within the traffic, as the network traffic is likely encrypted.

%\noindpar{Defining Privacy Leaks.} Define this! If possible?

\subsection{Design Goals} 
\sysname must make the underlying behavior of the devices in the home apparent, and enable forms of fine-grained (informed) control of the leakage of sensitive information for the user.
Towards this end, and addressing the challenges described in Section~\ref{ss:challenges}, we are guided by the following design goals.

(1) \textbf{Handheld and Mobile.} Smart devices are scattered throughout the home. Phone adoption is nearly universal. Using a phone as a window into the information world gives context and a sense of place. The phone form factor increases the likelihood of adoption and allows for inspection on the go; users can trigger or interact with devices and easily watch the movement of data, instead of having to return to the desktop.

(2) \textbf{Real-time.} Seeing statistics after the fact, as in most systems, is not as impactful or helpful when developing a model of how devices operate. Moreover, real-time analysis enables experimentation, providing users with a mechanism for exploring limitless scenarios and quickly associating triggers with outcomes.

(3) \textbf{Infrastructure/Hardware Free.} Many other methods require custom hardware. This increases cost and raises the barrier to entry. We hope to enable anyone, especially those that may have limited autonomy over infrastructure (i.e. renters, low-resourced communities) to be able to inspect the devices put in their living space.

(4)  \textbf{Intuitive Controls.} Complex mechanisms to control or limit the flow of privacy are not interpretable by users, and are possibly frustrating. Configuring a firewall is not a task most people would enjoy. Straightforward controls, with visible results, once those controls are put in place, are essential for users to trust the capability of the system.

(5) \textbf{Educational.} The ever-changing landscape of devices and the security/privacy arms is impossible to keep up with for privacy tools. Assisting users in understanding what makes certain devices leakier (e.g., always-on microphone) is essential.

To realize these design goals, we build the \sysname app---i.e., the handheld form factor---and \inspector as a two-part architecture working in tandem. Both systems must be running on the same local area network. \inspector, running on a computer, captures and analyzes network traffic between smart devices on the network and the Internet. \inspector acts as a server and provides the above information over an HTTP REST API. The \sysname app, acting as a client, regularly polls the REST API and presents the analysis as an AR overlay to users.

In the following sections, we detail the pieces of the system and how they interact to enable understanding and control of smart devices in the home.
In Section~\ref{ss:nettraffic} we discuss the \inspector and its role in capturing and curating privacy leak information; in Section~\ref{ss:visual} we describe the phone app design; in Section~\ref{ss:control} we detail the mechanisms we use for controlling devices on a schedule, and finally, in Section~\ref{ss:usecases} we describe a few ways to use \sysname.

% \begin{figure}[htp]
%     \centering
%     \includegraphics[width=8cm]{Figures/app_design.jpg}
% \end{figure}

\subsection{Low Burden Home Network Traffic Capture}\label{ss:nettraffic}

To use the \sysname app, the user must also have \inspector running on a computer (macOS, Windows, or Linux) that is on the same local area network as the phone. For our study, we run \inspector on a Raspberry Pi 3 Model B that is connected to the lab's network via Ethernet. We based \inspector{}'s code on the open-source project, IoT Inspector~\cite{huang2020iot}, and made modifications according to our needs. In particular, whereas the original IoT Inspector constantly sends captured traffic's metadata to the researchers' servers, \inspector runs without the Internet; it processes the captured traffic locally and exposes the traffic via a REST API. Furthermore, whereas the original IoT Inspector runs on users' computers and visualizes the traffic in a browser-based dashboard, \inspector uses an AR-based app, Privacy Plumber, to visualize the network traffic; the mobile app reads the processed traffic through the abovementioned REST API and presents the results as an AR overlay. 

Once running, \inspector automatically discovers IoT devices on the network, captures their network traffic via ARP spoofing, produces traffic statistics (e.g., bandwidth usage and identifying advertising and tracking services) over a local HTTP REST API, and blocks select devices (if desired by the user). We explain each of these features below.

\noindpar{Discovering IoT devices.} Upon launch, \inspector automatically broadcasts ARP packets to the local area network and discovers active devices. To identify IoT devices, Huang \textit{et al}.~\cite{huang2020iot} describe an algorithm that infers the likely identities of IoT devices based on MAC OUI (i.e., Organizationally Unique Identifier, basically the first three octets of a MAC address), DNS, and UPnP messages. For the prototype in this paper, we only use the MAC OUI. Within the code of \inspector, we have already hard-coded the mapping between OUIs and names of five IoT devices in our lab (which we can find out beforehand). In this way, \inspector can instantaneously identify the IoT device in our lab without relying on the device identification algorithm in Huang \textit{et al}.~\cite{huang2020iot}.

\noindpar{Capturing network traffic.} Once \inspector identifies a known IoT device on the lab's network, it automatically starts intercepting network traffic between the device and the Internet via ARP spoofing, a technique used in the original IoT Inspector implementation and which incurs an overhead of 3.4 Kbps, given that we have five IoT devices in the lab~\cite{huang2020iot}.\footnote{Per Huang \textit{et al}.~\cite{huang2020iot}, our setup includes $N = 5$ devices. It follows that $N(N+1)=30$ spoofed ARP packets are sent every two seconds. As each ARP packet has 28 bytes, the overhead is $28 \times 30 / 2 * 8 = 3,360$ Bits/second or 3.4 Kbps.}

\noindpar{Obtaining traffic statistics.} All traffic to and from IoT devices in our lab is redirected through \inspector. In doing so, \inspector is able to obtain statistics about the network traffic for every device, including the device's MAC address (from which to extract the OUI and determine the device's identity based on our hard-coded mapping); the number and size of packets (from which to infer the bandwidth usage); the remote IP addresses, DNS requests and responses, and the Server Name Indication field within TLS packets (from which to infer the remote hostname and whether the hostname is associated with a known advertising and tracking company, based on the Disconnect block list~\cite{DisconnectList}. \inspector presents all these statistics and information via an HTTP REST API that the \sysname app can access over the local area network. For example, if the computer running \inspector has a local IP address of $I_i$, then the \sysname app (on the same local network) can access the traffic information via \textit{http://[$I_i$]/get\_traffic}.

\noindpar{Phone Application: App Implementation.}
The \sysname mobile app was implemented in Unity using C\# and is cross-platform, tested on Android and iPhone. The app works by communicating with \inspector via HTTP GET requests, as described in the previous paragraph, to obtain JSON-encoded information about the devices on the network and their traffic. Parsing these JSON objects, the app visualizes the information as charts and text on the AR display (e.g., Figure~\ref{fig:live}). The app also shows an interface where users could block an IoT device's traffic, e.g., Figure~\ref{fig:control}. Once the user confirms, the app sends the corresponding request to \inspector via the HTTP REST API, and \inspector would subsequently block the device by jamming the device with corrupt ARP packets.

\subsection{User Control of Privacy Leaks from a Phone}\label{ss:control}
With \sysname we also want to help the user feel more empowered by allowing them to  take control of their devices with the ability to block device traffic. Users can manually block or allow device traffic indefinitely, or they can set rules to govern when they want their device to be on or off and for how long (Figure \ref{fig:control}). Users are also given the option to physically power off their device altogether. In this way, \sysname provides a closed-loop system where users can analyze the information flow out of a given device, then immediately apply direct control over that device in response and receive immediate feedback via the traffic view.

% \noindpar{Control Flow: Interaction between IoT Network Analyzer and App.}
%[Flow chart of how users can control / turn off devices using the app, which talks to the NTM, which turns the valve on devices.]
To illustrate how a user might control an IoT device's traffic, let us say that a user feels uncomfortable with an IoT device communicating with the Internet. The user can use the \sysname app to block Internet access on the device. As shown in Figure \ref{fig:control}, the user can click ``Block Traffic'' on the app to indefinitely block the device, or specify when to block and unblock the device. Moreover, the app sends an HTTP request to \inspector, using the REST API \footnote{http://[$I_i$]/block/[device\_id]/[block\_time]/[unblock\_time]} (where $I_i$ is the IP address of the running instance of \inspector). During the period of blocking, \inspector jams the communication of the device by using a corrupt source MAC address in the spoofed ARP packets (as described in Section~\ref{ss:nettraffic}). \inspector stops this process at \textit{[unblock\_time]}, upon which \inspector sends out spoofed ARP packets without the corrupt source MAC address. This gives users the ability to control the times of day when they want their devices to be on or off. 

Privacy Plumber's software-based device blocking offers several advantages over simply turning off or disconnecting a device. First, users do not need physical access to the device; for instance, many surveillance cameras are mounted on ceilings and are difficult to power off. Second, users can temporarily disable a device when they are feeling uncomfortable, e.g., blocking Amazon Echo for an hour during a sensitive phone call or conversation, through Privacy Plumber. Such temporary blocking is difficult to achieve through Echo's app (no such feature) or manually (e.g., the user has to remind themselves to re-connect Echo again). Third, though not currently implemented, Privacy Plumber, with the help of \inspector, can block based on the context (i.e., future work). For example, when \inspector detects the presence of a user's phone on the network (e.g., by checking if the phone responds to periodic ARP requests), \inspector automatically blocks all indoor cameras; when the phone leaves the network (e.g., when the user is out), \inspector could automatically unblock all indoor cameras.

\noindpar{Technical Mechanism for Blocking Devices.} A major difference with respect to IoT Inspector's original implementation is that we have added the capability of blocking devices in \inspector. Using the HTTP REST API \footnote{http://[$I_i$]/block/[device\_id]/[block\_time]}, the \sysname app can request \inspector to block a certain device at a particular time (for instance, because the user does not want the device to be communicating to the Internet). Upon receiving this request, \inspector jams the network communication of the device by sending it spoofed ARP packets with corrupt MAC addresses.

To illustrate this process, let us assume that the computer running \inspector has a MAC address $M_i$ and IP address $I_i$. Let us also assume that \inspector is about to intercept the communication from the gateway (with MAC address $M_g$ and IP address $I_g$) to a particular device (with MAC $M_d$ and IP $I_d$) without blocking. To do so, every two seconds, \inspector sends an ARP packet to the device, such that the ARP packet has a source MAC of $M_i$ and a source IP of $I_g$, along with a destination MAC of $M_d$ and a destination IP of $I_d$. In contrast, let us assume that \inspector is to block the device. It sends the same ARP packet to the device, except that the ARP packet's source MAC is a series of random numbers (instead of $M_i$) that represent an unreachable MAC address on the local area network.

\subsection{Visualizing and Understanding Traffic in Real-Time}\label{ss:visual}
% For the longest time, network traffic has been invisible. Users have never really been able to see how their information is being used, let alone be aware of how their information is being disclosed to third parties. 
One of the goals of \sysname is to show users contextualized network activities of IoT devices to help them pinpoint the potential privacy risks. In this section,  we discuss how \sysname utilizes Augmented Reality to help users contextually visualize their devices' network traffic information in real-time, provide a chart of network traffic in real-time, and provide links to other research in which the privacy concerns of the inspected device have been studied (including home behavior inference, sleeping behaviors, and personal data). Lastly, users are able to send feedback and bug reports.

\noindpar{Use of Augmented Reality.}
The use of AR visualization makes the interaction with the device the user is inspecting more tangible and contextual. While IoT Inspector~\cite{huang2020iot} and \inspector are text-only data-driven analyzers that can only be accessed using browser HTTP requests, \sysname is a fully-fledged interactive application due to the utilization of AR. By pointing their camera at the device being inspected, the user can see, in their environment, the traffic coming out of the device that they are physically inspecting. Users can interact with their devices and receive immediate feedback about data output and communication with advertisers. Combined with manual device control, this is intended to help the user feel informed and in control of the IoT devices that physically surround them, similar to the use of a TV remote control. 

\noindpar{Learning About Privacy Threats.}
We aim to educate and inform users on how their IoT devices expose their network traffic information to third parties. In Figures~\ref{fig:education} and \ref{fig:literatureapp}, the app shows icons surrounding the IoT device. When any of these icons are pressed, they provide links to other research materials---which we have manually curated in advance---where the privacy concerns of the inspected device have been studied. Depending on the device, \sysname provides the following categories of potential privacy violations represented by icons: 

\begin{itemize}
\itemsep 0.1em 
    \item Location: Your physical location either roughly (your address) or fine-grained (room you are in).
    \item Activity: Your physical activity such as walking, talking, sleeping.
    \item Screen: Your online activity, such as when you browse videos on YouTube or surf the web.
    \item Identity: Attributes that can identify you such as your face or voice.
    \item Shopping: Data on your usage of money or products.
    \item Health: Can infer different health markers without consent (heart rate, breathing, and others).
\end{itemize}

%Database of Detectable Behaviors and Relevant Devices
%We did a bunch of work to get the scary papers on behaviors extractable from smart home devices (Alexa detects your heart rate and so on).
%\josiah{Talk about the information displays here? All the different potential things that can happen, we can also draw from Mozilla's list of devices as well as our own lit.}

\subsection{\sysname Example Use Cases}\label{ss:usecases}

In this section, we illustrate two example use cases of the \sysname app. We focus on the ability of \sysname to enable experimentation and the usefulness of a real-time inspector. We will describe the users' reactions in Section 4.3.

\noindpar{Example 1: Is Echo Always Listening?} 

\noindent A user may use the \sysname app to correlate network activities on an Amazon Echo device with the user's interactions---or the lack thereof---with it. While pointing the AR camera at the device, the user could invoke a voice command, such as ``Alexa, what is the weather'', while observing the device's bandwidth usage graph on the \sysname app. Afterward, the user may physically press the mute button on Echo, repeat the same voice command, and observe the bandwidth usage graph on the app.

\noindpar{Example 2: What is this App on My Smart Fridge?} 

\noindent Many smart fridges have built-in touch-screen panels. For example, the Samsung Smart Fridge has a tablet-like touch-screen panel to control various settings of the fridge (such as temperature). The panel also allows users to access various built-in apps, such as checking recipes or ordering ingredients online. A user who is concerned with the privacy of such apps may point the AR camera at the fridge, interact with an app, and observe the advertising and tracking services counter on the app. This counter shows, in real-time, the total number of advertising and tracking services that the fridge has communicated with, based on the Disconnect block list~\cite{DisconnectList}.

\section{Pilot User Study}

To test how users react to \sysname and inform its future iteration,  we conducted a pilot study with 6 participants to experiment with, understand, and control the potential privacy violations of IoT devices. It should be noted that the pilot study would be best conducted in participants' homes. However, due to University research restrictions, the COVID-19 pandemic has made it difficult for us to recruit real users, distribute hardware (e.g., phones powerful enough for AR and Raspberry Pi's for running \inspector), and conduct a free-living study.

We conducted a one-day controlled lab study in our IoT Lab with 6 participants. Participants were invited to use the \sysname app while interacting with several IoT devices in the lab, including Samsung Smart Fridge, Amazon Echo, Google Home, Samsung Smart TV, and Google's Nest Cam. Our goal is to assess whether using augmented reality to display network traffic (i.e., by using \sysname) influenced the participants' awareness of privacy and changed their behaviors. 

% Our study shows that \sysname increased participants' understanding of network traffic, changed participants' awareness, raised privacy concerns, and educated them on how to better manage and take control of how IoT devices handle their information. 
In the following sections, we present the details of the pilot study and discuss some highlights in the results as well as lessons learned to inform the next iterative of \sysname.
% discuss the procedure of the study, the study results, and provide information on network overhead and \sysname's battery lifetime.

%\josiah{Make sure these map directly to contributions and or goals in intro!!}
%\begin{enumerate}
  %  \item Do we enable understanding of network traffic and the behaviors that could be predicted from it? Is the understanding accurate? How precise / granular? What are the confounders. [quantitative approach]
  %  \item How effective are the controls we implemented? And resource / packet contention OK? [i.e. do we impact network performance in a negative way with our control ability]
   % \item What are the resource requirements for the system and are they manageable? (i.e. does this kill your battery in ten minutes). WE argue that the high cost is worth the information, and its not something you are running all the time.
%\item Does this improve privacy literacy? Integrate this with the user study and talk to Marshini or Ada.
%\end{enumerate}

\subsection{Participants Recruitment and Demographics} 
% Due to the COVID-19 pandemic, we were not able to recruit participants from the general population. Instead, w
We recruited 6 graduate students from our institution through our university mailing list. We did so rather than recruiting from a broader population sample because of the constraints our university implemented during the COVID-19 pandemic (i.e., external members were not permitted to enter our buildings). Our sample consisted of four males and two females. Three of the participants were between the ages of 18-24, two participants were between the ages of 25-34, and one participant was between the ages of 35-44. 
% They come from a variety of majors (e.g., ****) and different levels of study (e.g., ** undergraduate students, ** graduate students).

% \yaxing{@danny, do you have details of the demographics?} \danny{No more details}

\subsection{Study Procedure and Data Collection}

For safety reasons and to implement social distancing procedures, only two people were allowed in the IoT Lab during the study. Aside from the participant, one of the co-authors in this paper served as the research coordinator. They were present throughout the user study to help guide the participants or troubleshoot any technical difficulties that could arise during the study procedure. 

Before the study began, each participant filled out a background pre-survey on a computer in the IoT lab. We asked questions about their demographics, how technically savvy they are, their smart device experiences, their general understanding of privacy, and their concerns about their information being exposed to third parties.

After completing the survey, our research coordinator handed each participant a script and an Android mobile phone that had \sysname installed. Following the script, each participant opened the \sysname app, kept it running in the foreground, and interacted with one IoT device at a time. Regardless of the IoT device, each interaction consists of the following steps, as prescribed in the script:

\begin{enumerate}
    \item Using the \sysname app, the participant scanned the QR code that we had placed on the IoT device. The QR code encodes the device's MAC address, device name, and manufacturer. Based on the QR code, \sysname shows the corresponding device's AR model on the screen.
    \item The participant used the app to inspect the device's traffic, while not doing anything to the device.
    \item The participant interacted with the device (which we will describe in detail). During the interactions, the participants observed the network traffic graph on the app. 
    \item Using the app, the participant clicked on any of the icons surrounding the AR model of the device and read the educational material.
    % \item Using the app, the participant would give an emoji or sequence of emoji's describing their feelings and emotions on this device and on what it shared with third parties. 

\end{enumerate}

After interacting with all the IoT devices, participants returned the phone to the research coordinator and responded to a post-survey that asked the same questions as in the pre-survey, along with a usability survey. We discuss the results in more depth in Sections~\ref{sec:results:awareness} and \ref{sec:results:usability}. We also include our pre- and post-surveys in the Appendix. %\danny{Make sure the surveys are in the Appendix.}

\begin{figure}[t]
    \centering
    \includegraphics[width=0.7\columnwidth]{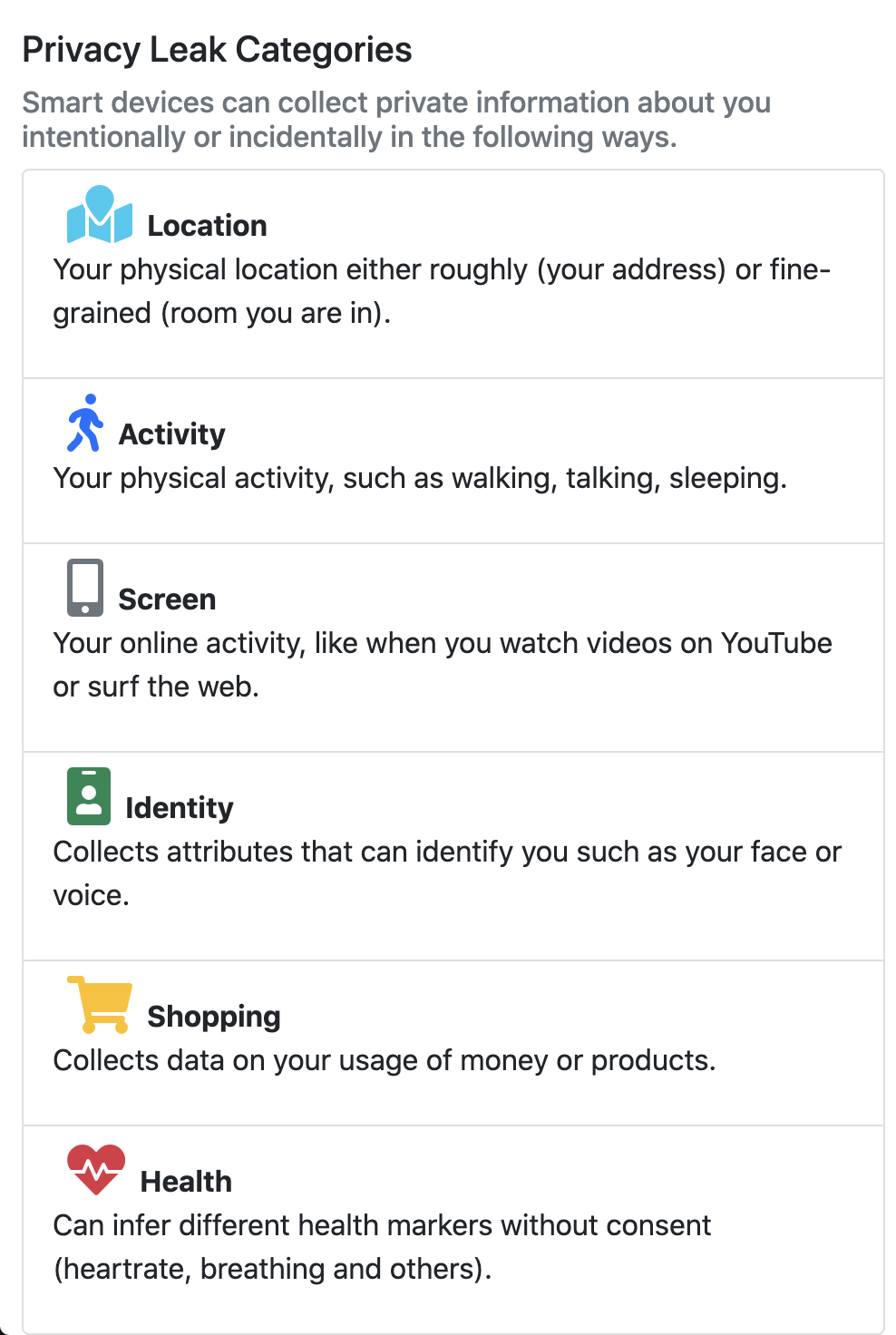}
    \caption{This screen on the phone application describes the different categories of privacy leaks that different devices have, based on a database that we manually curated in advance.}
    \label{fig:literatureapp}
\end{figure}

Below, we describe each participant's scripted interactions with each device---i.e., showing Step 3 in detail. During the interactions with the devices, users can access the educational content which is summarized from Mozilla's ``privacy not included'' handout~\cite{mozilla-privacy} and academic literature. Each device is described by the categories of privacy exposure they create, those categories are shown in Figure~\ref{fig:literatureapp}.

\begin{figure*}[t]
    \centering
    \includegraphics[width=0.8\textwidth]{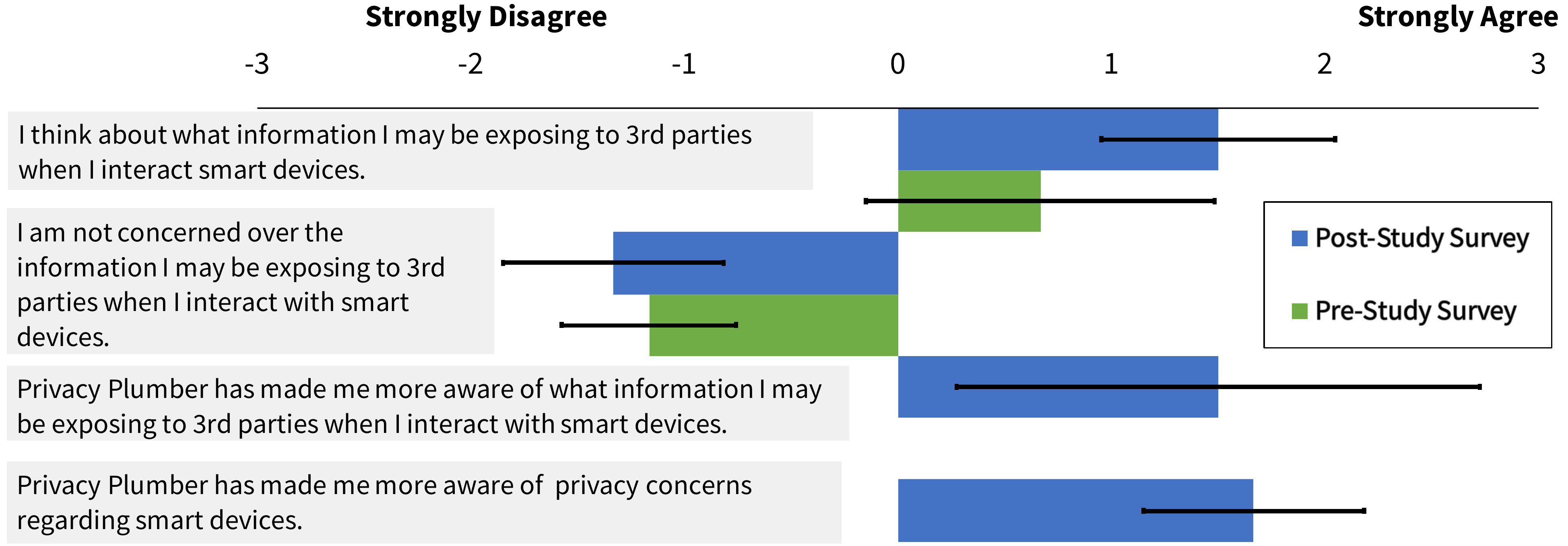}
    \caption{Representation of participants' average agreement ratings relating to statements about information being exposed to third parties and privacy concerns caused by interacting with IoT devices. Participants rated the first two statements before and after the study, while the last two statements were rated at the end of the study. The results show that after the study, participants displayed an increase in awareness and concern about how their information is being handled when interacting with IoT devices.}
    \label{fig:Graphs}
\end{figure*}

\noindpar{Samsung Smart Fridge.} The fridge has a built-in touchscreen on the door. Through the touchscreen, users have the ability to interact with several built-in apps, such as managing the shopping list, checking what is inside the fridge, and searching for recipes online. Users can also interact with the touchscreen using voice commands, using the trigger word, ``Bixby.'' 

Per the script, the research coordinator instructed the participant to follow the following three tasks. (i)~Once the participant scanned the QR code of the smart fridge, they said the voice command, ``Hey Bixby, do we have mangoes?'' Bixby, the fridge's voice assistant, would say ``no,''. The participant then said, ``Hey Bixby, can you add mangos to my shopping list?" Immediately, the participant looked at the \sysname app and observed the network traffic emitting from the fridge for about 30 seconds. (ii)~The participant said, ``Hey Bixby, find me a Ramen recipe.'' The recipe app popped up on the touchscreen. Using the finger, the participant browsed through the available recipes on the screen, while observing the network traffic on \sysname for 30 seconds. (iii)~The participant opened the fridge door and then closed it. Once again, they inspected the fridge's network traffic through the \sysname app for 30 seconds. 

\noindpar{Amazon Echo.} Interactions with Echo consists of the following 3 tasks. (i)~The participant said  the voice command,  ``Alexa, play a thunderstorm sound." Immediately, the participant observed the network traffic on the app for 30 seconds. (ii)~The participant physically pressed the ``mute'' button on the Echo and watch the device's network traffic for 15 seconds. (iii)~The participant said the same voice command as in Task (i) and observed the traffic in the app. 

\noindpar{Google Home.} The participant said a voice command, ``Hey Google, what was the final score in the Super Bowl last year?'' The participant immediately started observing the network traffic on the app for 30 seconds. 

\noindpar{Samsung Smart TV.} The participant used the TV's remote control to navigate to the Bloomberg app on the home screen. They then started streaming a live video on the Bloomberg app for one minute while they observed the network traffic with the \sysname app. 

\noindpar{Nest Cam.} Interactions with the camera consists of the following 2 tasks. (i)~The participant walked into the field of view of the camera and stay there for five seconds, walked out of the camera's field of view, and observed inspect the network traffic with the \sysname app. They repeated this task as many times as they liked. (ii)~The participant blocked the network traffic to and from the camera using the built-in feature on the \sysname app. The participant observed the network traffic for 10 seconds, walked in front of the camera's field of view, waited for another ten seconds, and unblocked the device using the \sysname functionality.

\subsection{Analysis of Pre-Study and Post-Study Surveys}
\label{sec:results:awareness}

We asked each participant to complete two surveys: (i) a pre-Study Survey that they filled out on a dedicated computer at the beginning of the study, i.e., before the participants interacted with the \sysname app or the IoT devices; 
% (ii) an Emoji Survey that the participants filled out on the \sysname app during the interactions with each of the IoT devices; and 
(ii) a post-Study Survey that the participants filled out on the dedicated computer after interacting with all the five IoT devices. We present the results below.

% In this section we analyze the results from all three surveys.  Results from the Pre- and Post-Study Surveys demonstrate that \sysname improved the participants' awareness of the privacy and security risks of IoT devices, and that using \sysname helped participants understand the network traffic that was generated when they interacted with the IoT devices used in the study.  
% In addition, results from the Emoji Survey reveal that many participants felt sad and disappointed.

% \noindpar{Analysis of the Pre- and Post-Study Surveys.}
In Figure~\ref{fig:Graphs} we present the participants' agreement rating responses for two statements that were asked in the pre-study survey and post-study survey. We observe that for those two statements participants seemed less concerned by how their information is exposed to third parties when they interact with IoT devices before they performed the activities in the study. After participants completed the study, they were more aware and concerned about how their information was disclosed to third parties. The last two statements of Figure~\ref{fig:Graphs} were only given in the post-study survey, which asked participants to rate whether \sysname was useful in helping them become more aware of privacy concerns and how their information is being shared with third parties. On average, participants somewhat agreed that \sysname helped raise their awareness and privacy concerns. Participants found that \sysname was helpful in that it helped them visualize what information was being shared.

% Please add the following required packages to your document preamble:
% \usepackage{booktabs}
% \usepackage[table,xcdraw]{xcolor}
% If you use beamer only pass "xcolor=table" option, i.e. \documentclass[xcolor=table]{beamer}
% \usepackage[normalem]{ulem}
% \useunder{\uline}{\ul}{}
\begin{table*}[t]
\small
\begin{tabular}{@{}lllllllllll@{}}
\toprule
\textbf{Survey Question} &
  \multicolumn{2}{c}{\textbf{Smart Fridge}} &
  \multicolumn{2}{c}{\textbf{Amazon Echo}} &
  \multicolumn{2}{c}{\textbf{Google Home}} &
  \multicolumn{2}{c}{\textbf{Smart TV}} &
  \multicolumn{2}{c}{\textbf{Nest Cam}} \\ \midrule
 &
  \multicolumn{1}{l|}{\textit{pre}} &
  \cellcolor[HTML]{EFEFEF}\textit{post} &
  \multicolumn{1}{l|}{\textit{pre}} &
  \cellcolor[HTML]{EFEFEF}\textit{post} &
  \multicolumn{1}{l|}{\textit{pre}} &
  \cellcolor[HTML]{EFEFEF}\textit{post} &
  \multicolumn{1}{l|}{\textit{pre}} &
  \cellcolor[HTML]{EFEFEF}\textit{post} &
  \multicolumn{1}{l|}{\textit{pre}} &
  \cellcolor[HTML]{EFEFEF}\textit{post} \\
\begin{tabular}[c]{@{}l@{}}I think this device could be (or is) useful or\\ valuable to my daily life and routine.\end{tabular} &
  \multicolumn{1}{l|}{3} &
  \cellcolor[HTML]{EFEFEF}3.17 &
  \multicolumn{1}{l|}{2.86} &
  \cellcolor[HTML]{EFEFEF}2.5 &
  \multicolumn{1}{l|}{2.71} &
  \cellcolor[HTML]{EFEFEF}2.5 &
  \multicolumn{1}{l|}{2.43} &
  \cellcolor[HTML]{EFEFEF}2.33 &
  \multicolumn{1}{l|}{2.71} &
  \cellcolor[HTML]{EFEFEF}3 \\ \cmidrule(r){1-1}
\begin{tabular}[c]{@{}l@{}}I am comfortable having this device in\\ my house and always on.\end{tabular} &
  \multicolumn{1}{l|}{2.29} &
  \cellcolor[HTML]{EFEFEF}3.5 &
  \multicolumn{1}{l|}{3.86} &
  \cellcolor[HTML]{EFEFEF}4.17 &
  \multicolumn{1}{l|}{3.86} &
  \cellcolor[HTML]{EFEFEF}4.17 &
  \multicolumn{1}{l|}{2.29} &
  \cellcolor[HTML]{EFEFEF}3.5 &
  \multicolumn{1}{l|}{3.43} &
  \cellcolor[HTML]{EFEFEF}4.17 \\ \cmidrule(r){1-1}
\begin{tabular}[c]{@{}l@{}}I am comfortable having this device in\\ my house if I can automatically control\\ when it is on, or off.\end{tabular} &
  \multicolumn{1}{l|}{1.29} &
  \cellcolor[HTML]{EFEFEF}2.17 &
  \multicolumn{1}{l|}{2.29} &
  \cellcolor[HTML]{EFEFEF}2.5 &
  \multicolumn{1}{l|}{2} &
  \cellcolor[HTML]{EFEFEF}2.33 &
  \multicolumn{1}{l|}{1.14} &
  \cellcolor[HTML]{EFEFEF}2 &
  \multicolumn{1}{l|}{2.29} &
  \cellcolor[HTML]{EFEFEF}2.83 \\ \midrule
\textit{} &
   &
   &
  \multicolumn{1}{l|}{} &
  \multicolumn{7}{c|}{Strongly Disagree (5) to Strongly Agree (1)} \\ \cmidrule(l){5-11} 
\end{tabular}
\caption{Results of the survey on user awareness and comfort with smart devices, before and after using Privacy Plumber to inspect those devices. Scores are listed for both pre- and post-study surveys for each device. The higher the scores, the more strongly the participant disagreed with the survey question statement.}
\label{tab:awareness}

\end{table*}

Additionally, we discuss the results of participants' responses with the IoT devices before and after the study. We show that after the study participants felt less safe with how IoT devices handle their data. Participants were presented with three statements and were asked to rate whether they agree or disagree with these statements on a scale of one to five, where a 1 meant they strongly agree and a 5 represents a strongly disagree rating. Table~\ref{tab:awareness} demonstrates the average change in attitudes participants had before the study and after the study. We note that before the study, on average participants neither agreed nor disagreed with the statements presented in Table~\ref{tab:awareness}. After completing the study, the average rating agreement score increased to ``somewhat agree'' on the last two statements on all IoT devices. The exception was in the first statement, the scores for the Amazon Echo and Google Home. This indicates that after using \sysname in the study, participants felt that the Amazon Echo and Google would still find it useful to use in their households.

We also observe that the Smart Fridge, Smart TV, and the Nest cam had the most significant change in attitude. 
% Danny: I removed all instances of "creepy" as that's not precise.
%Participants felt these devices were the ``creepiest.'' 
We gathered a few quotes from participants in which they describe how they felt about interacting with these IoT devices and using \sysname to inspect their network traffic:

\begin{displayquote}
\textit{IoT devices provide more information to third parties than people thought. I think apps like Privacy Plumber can help people to make better decisions when using IoT devices} --- \textbf{(P1)}
\end{displayquote}

\begin{displayquote}
\textit{Cool to see when and how much traffic each device sends at any given moment!} --- \textbf{(P5)}
\end{displayquote}

\begin{displayquote}
\textit{I think the app does make me more aware about how the traffic is associated with the behavior of the device. Having some control over the traffic is nice. That being said, if I do have the device in my home, I probably would like to use it, and in that case, I have to allow traffic, which I have no control about what could pass or could not pass. In that sense, I can only accept certain privacy risks.} --- \textbf{(P2)}
\end{displayquote}

\begin{displayquote}
\textit{It was interesting to see the potential privacy leaks shown next to the device. Some leaks/ privacy implications were surprising. Liked the ability to allow/block traffic, it was also cool to see the real-time traffic including communication with third-party advertisers. Liked the app interface.} ---\textbf{(P6)}
\end{displayquote}

These quotes, along with results from Figure~\ref{fig:Graphs} and Table~\ref{tab:awareness}, suggest that \sysname helped participants  understand the network traffic, increased their awareness of potential privacy violations, and helped them make more informed decisions on how to handle IoT devices.

\subsection{Analysis of the Usability Survey}
\label{sec:results:usability}

At the end of the study, each participant completed the usability survey. Overall, most participants indicated that they would use \sysname in their home network, found it easy to use and user-friendly, and agreed that most people would learn to use \sysname quickly. We summarize the results below:

\begin{itemize}
    
    \item When asked if they would use the \sysname mobile app to inspect the data the IoT devices in their homes, two participants said they strongly agreed with the statement and four participants said they somewhat agreed to use \sysname. 

    \item When participants were asked if they found \sysname easy to use, four of them somewhat agreed, one participant strongly agreed, and one participant somewhat disagreed. 
    
    \item When presented the statement ``I imagine that most people would learn to use Privacy Plumber very quickly'', the responses were across the board spectrum.  Three participants rated that comment as strongly agreed, one participant rated the statement with a somewhat agree, one participant responded that they felt neither agreed or disagreed with the statement, and one participant somewhat disagreed.

    \item When participants were asked to rate the overall user-friendliness's of \sysname, four participants rated the \sysname app as good and two participants rated \sysname as fair. 
\end{itemize}

We gave participants an open-ended question if they would improve the usability of \sysname, and if so, how. We show their responses in Appendix~\ref{sec:FreeResponseUsabilitySurvey}. All in all, participants seemed to respond somewhat positively towards \sysname. It shows that \sysname may have the potential to be distributed to the general public after further studies. We hope to build off our current platform and implement the suggestions our participants gave us in future work.

% \danny{I recommend our base case be 2D infographics (similar to Lorrie Cranor's privacy labels). We show that our 3D/4D AR approach is more effective to communicate privacy risks than traditional 2D approach, and that users will more likley have an accurate mental model of the privacy risks.}

\subsection{Performance: System Overhead and Battery Life Impact}

\noindpar{Network Overhead.} 
\inspector intercepts the network traffic of select IoT devices via ARP spoofing, a technique that could introduce network overhead especially to the targeted IoT devices. This overhead comes from two sources. First, the spoofed ARP packets consume extra bandwidth, although the overhead is relatively small---i.e., less than 60 Kilobytes/second even if 50 IoT devices are under ARP spoofing~\cite{huang2020iot}). The second source of overhead comes from the Raspberry Pi 3 Model B, where we run \inspector in the lab. The Raspberry Pi is connected to the lab's network via Ethernet. For all IoT devices to which \inspector sends spoofed ARP packets, all inbound (i.e., download) and outbound (i.e., upload) traffic to and from the IoT devices has to first go through the Raspberry Pi before \inspector forwards the traffic to the targeted device and to the Internet respectively. Effectively, the Raspberry Pi introduces a bottleneck for the ARP-spoofed devices.

To measure the overhead as a result of the Raspberry Pi bottleneck, we conduct the following experiment. We install the Ookla Speed Test app on an Android phone that is connected to the the lab's WiFi network. We have the Ookla app run 15 back-to-back speed tests, which measure the inbound and outbound traffic rates with respect to a server in our city, as well as the latency of packets. Using the same setup, we repeat the same experiment, except that we have \inspector inspect the phone's traffic via ARP spoofing. 

We find significant overhead as a result of \inspector. Without ARP spoofing, the app achieves, on average, an inbound rate of $293.6\pm15.4$ Mbps, an outbound rate of $94.1\pm0.2$ Mbps, and a latency of $5.7\pm0.5$ milliseconds. With ARP spoofing by \inspector, the app achieves, on average, an inbound rate of $41.4\pm74.6$ Mbps, an outbound rate of $72.8\pm14.1$ Mbps, and a latency of $5.9\pm0.5$ milliseconds. Compared with the case without ARP spoofing, \inspector reduces the inbound rate by 85.9\% and outbound rate by 22.6\%, while increasing the latency by 3.5\%. 

Despite the seemingly significant reduction in bandwidth, we argue that \inspector is unlikely to degrade usability, as the network analyzer is not always running (only when inspecting, or blocking a specific device). Additionally, the overhead can be reduced with improved hardware. According to Netflix, 25 Mbps of inbound rate is sufficient to stream Ultra HD contents~\cite{NetflixHDSpeed}. A user who inspects a smart TV using \inspector is likely to enjoy Ultra HD streaming given the reduced inbound rate of $41.4\pm74.6$ Mbps under ARP spoofing. If a user desires to reduce the network overhead, the user could upgrade the computer that runs \inspector, as Raspberry Pi 3 is anecdotally known for its poor networking performance~\cite{RaspberryPiPoorNetworkingPerformance1,RaspberryPiPoorNetworkingPerformance2}. Possible upgrade option could include a computer---or ODroid if the user needs the compact form factor~\cite{ODroid}---that is shipped with a fast CPU and a Gigabit Ethernet card.

\noindpar{Battery Lifetime.} 
We used AccuBattery on android, to try to understand the energy cost. This does not hold across phones, so we compare the energy cost against YouTube and TikTok for ten minutes of streaming video.
With all the background application killed, 10 minutes of Privacy Plumber impacts 3.98\% (159mAh) of the battery lifetime, while YouTube costs 2.63\% (105mAh) and TikTok costs 3.9\% (156mAh).
Privacy Plumber is only meant for point inspection and short usage to analyze new devices in the home, or experiment with different setups, so it should not impact battery lifetime too much since it is not always on.
Moreover, the battery lifetime cost is similar to that of streaming videos online, a normal function, therefore users should not expect significant battery lifetime loss due to usage of \sysname.

% If successful, VisIot can improve the limitations of IoT inspector by providing a tool that will enable users to be contextually aware of third party websites accessing their data in real time by using the AR application.  There is no need to install software on your desktop or laptop. It would be as easy as downloading an app on an iphone or ipad. As a result, users who are not technologically savvy, can easily point the IoT device of interest with their ios device and observe the data exhaust of that IoT device in real time. VisIoT can therefore be easily accessible to the general public to use and allow them to take control of their security and privacy. 
\section{Discussion on Limitations and Future Work}\label{sec:discussion}

% \noindpar{Bad Mental Models.} \sysname is able to in the moment associate network traffic, trackers, and other data visually with user actions or interactions with a device. This can be good for building mental models, however, there is a risk the mental model may be inaccurate...

% A mute button affords a certain function and user expectations

\noindpar{Comparing users' mental models against actual contents of IoT network traffic.} Our results show that users' mental model of how IoT devices communicate with the Internet may be inconsistent with how devices \textit{appear} to behave, but it is unclear whether this mental model is consistent with the \textit{actual} contents of the communication. For example, two participants in our study did not expect network traffic from Amazon Echo when the device's microphone was on mute. Presumably, the participants expected Amazon not to send any audio data back to Amazon during mute. In this case, Echo's \textit{apparent} behavior was the communication with the Internet on mute; in contrast, whether Echo \textit{actually} sent out audio data was unknown. Our system did not extract the contents of the communication, which could be encrypted based on previous results~\cite{apthorpe2019keeping}.

Despite the encrypted contents, man-in-the-middling is possible (e.g., per Moghaddam \etc~\cite{mohajeri2019watching}). In future in-lab studies, we plan to modify \inspector to intercept and decrypt IoT traffic, assuming that devices do not validate certificates and/or do not use certificate pinning. We hope to extract the payload from some of the TLS connections, identify exactly what devices are sending to the Internet, and compare it against users' mental models.

\noindpar{Automated, contextualized blocking of devices.} The current prototype allows users to set a block/unblock schedule for IoT devices. Although this feature provides users with fine-grained control, it requires manual effort from the user both in setting what devices to block and when to block. 

We plan to augment this feature with automated device blocking based on contextualized information that \inspector already collects. For example, a user could create a rule on \inspector that would automatically block surveillance cameras if \inspector detects the presence of mobile phones (based on ARP and pings) in the home network (which could suggest that the residents are home); otherwise, it can unblock the cameras to capture, say, unauthorized entry into the property. As another example, let's say a user has an Amazon Echo and a smart TV in the living room. The user could create another rule that lets \inspector automatically block Amazon Echo if it detects active streaming traffic from the smart TV, as the user may not want Echo to capture any conversations while the family is watching TV in the living room. In short, by leveraging the IoT traffic that \inspector already collects, users could create automated, contextualized rules to block IoT devices from collecting sensitive data.

\noindpar{Deployment roadmap and challenges.} 
We plan to deploy the \sysname app and \inspector to real-world users at scale. Based on our current prototype, we plan to make the following modifications.

\textit{Operating system support}. Once deployed, our system will have the same two-component architecture, although we will expand the \sysname app to both iOS and Android (current prototype), and \inspector to all major non-mobile operating systems including macOS, Windows, and Linux (current prototype). This process will likely be straightforward, as we developed both components with cross-OS platforms (Unity for the app and pure Python for \inspector). 
    
\textit{Network-based device identification}. We will develop network-based device identification mechanisms to help users distinguish among their devices and identify the device(s) of interest. The current prototype identifies devices based on a hard-coded mapping between MAC OUIs and device names, because we already know the inventory of IoT devices in the lab. For real-world deployment, we will incorporate IoT Inspector's device identification algorithm~\cite{huang2020iot}, so that our system will dynamically infer device names based on the network signature, which includes not only OUIs, but also DNS queries, UPnP banners, mDNS names, and DHCP hostnames. \edited{We will also use information in the 802.11 frames to discover and locate devices~\cite{lumos}.}

\textit{Image-based device identification}. To complement the network-based approach, we will also develop image-based device identification mechanisms for the AR camera. Currently, the \sysname app identifies devices based on printed QR codes on or near select IoT devices, such that the QR codes encode the MAC addresses and the names of devices. For real-world deployment, we will use computer vision to train a model of common IoT device types, such as voice assistants, smart TVs, and surveillance cameras (where security and privacy issues are commonly found in the literature). This model will help the AR app recognize \textit{possible} IoT devices (e.g., ``likely a smart TV''). The app will then refine the recognition with the network-based device identification algorithm (e.g., ``whether the device is indeed a smart TV based on the network signatures'') and manual user input if necessary. Both the network- and image-based approaches will hopefully help the app identify IoT devices in real-world settings.

\noindpar{Expanded user study.} 
\edited{
The user study, as a pilot, has a small sample size and is limited to graduate students, who may be more inquisitive or technically-inclined than the general population. We hope to scale out the testing to a larger userbase, both in lab and in real homes, in future work. We will also compare the participants' changes in privacy awareness against other visualization tools (e.g., IoT Inspector~\cite{huang2020iot} and Aretha~\cite{seymour2020informing}). Finally, we will conduct in-depth studies on various ways to visualize privacy leaks in AR (e.g., icon overlays and animations).
}

\section{Summary}

This paper presented Privacy Plumber, an end-to-end system demonstrating how a general population of end users can \edited{potentially} have insight into the network traffic of smart home IoT devices, and how these users can control when these smart devices could communicate with the Internet with one click of a button.
Designed after the concept of a leak detector, Privacy Plumber is a phone app with a tethered desktop application---\inspector---that provides an inspect and correct interface supported by network traffic analysis (inspect) and automated and timed network traffic jamming (correct).

Privacy Plumber is the first real-world inspection and control system that can be deployed in any home without new hardware or router modifications. 
\edited{Using AR, the tool aims to help users model IoT device activities within the context of the physical environment and of user interactions (addressing challenges C1 and C3, per Section~\ref{ss:challenges}); 
it gives users the option to block IoT devices and control the privacy ``valve'' (C2); 
it provides users with an interface to visualize IoT device activities as users interact with devices (C4);
and it requires a modern AR-supported phone and computer, without any dedicated or specialized hardware (C5).
}

We evaluated Privacy Plumber inside an instrumented smart home space with a variety of devices not previously evaluated for any privacy-enhancing tool, including a smart fridge, a smart TV, voice assistants, and Internet-connected surveillance cameras.
We found that using Privacy Plumber improved users' awareness of potential privacy violations of devices and that the system was generally easy to use and afforded useful controls.
In the future, we hope tools like Privacy Plumber will give mechanisms back to the user for stymieing the flow of private information outside the home, especially as our homes and living spaces become smarter, often without our consent.

\section*{Acknowledgment}

\edited{This research is based upon work supported by the National Science Foundation under award numbers CNS-2219867, CNS-1739809, and CNS-1915847. Any opinions, findings,
and conclusions or recommendations expressed in this material are
those of the authors and do not necessarily reflect the views of the
National Science Foundation.
The research is also based on work supported by gifts from Consumer Reports and Meta.}

\bibliographystyle{plain}
\bibliography{refs}

\appendix

\section*{Survey Questions}

\edited{All questions require responses in Likert scales, ranging from ``Strongly Agree" (1) to ``Strongly Disagree" (5).}

\subsection{Pre-Study Survey Questions}

\begin{enumerate}
    \item When I am in a smart home, I think about what information I may be exposing to vendors, companies, and 3rd parties when I interact with or sit in the same space with smart devices in the home.

    \item I am not concerned about the information I may be exposing to 3rd parties when I interact with or sit in the same space as smart devices in a smart home.

    \item I think this device could be (or is) useful or valuable to my daily life and routine. 

    \begin{itemize}
        \item Smart Fridge
        \item Google Home
        \item Amazon Echo
        \item Smart TV
        \item Nest Cam
    \end{itemize}

    \item I am comfortable having this device in my house and always on.
    
    \begin{itemize}
        \item Smart Fridge
        \item Google Home
        \item Amazon Echo
        \item Smart TV
        \item Nest Cam
    \end{itemize}

\end{enumerate}

\subsection{Post-Study Survey Questions}

\begin{enumerate}
    \item When I am in a smart home, I think about what information I may be exposing to vendors, companies, and 3rd parties when I interact with or sit in the same space with smart devices in the home.

    \item I am not concerned about the information I may be exposing to 3rd parties when I interact with or sit in the same space as smart devices in a smart home.

    \item Privacy Plumber has made me more aware of what information I may be exposing to 3rd parties when I interact with smart devices in the home.

    \item I feel Privacy Plumber has made me more aware of privacy and security concerns surrounding IoT devices.

    \item I think this device could be (or is) useful or valuable to my daily life and routine. 

    \begin{itemize}
        \item Smart Fridge
        \item Google Home
        \item Amazon Echo
        \item Smart TV
        \item Nest Cam
    \end{itemize}

    \item I am comfortable having this device in my house and always on.
    
    \begin{itemize}
        \item Smart Fridge
        \item Google Home
        \item Amazon Echo
        \item Smart TV
        \item Nest Cam
    \end{itemize}

    \item Finally, please provide any other thoughts or observations from participating in this experiment with Privacy Plumber (open ended).

\end{enumerate}

\section*{Additional Responses from the Usability Survey}
\label{sec:FreeResponseUsabilitySurvey}

We gave participants an open-ended question if they would improve the usability if privacy plumber, if so how. We obtained the following responses from each participant.

    \textit{I would include more guidance or instructions in the app for first-time users.} \textbf{(P1)}
    
    \textit{I think the app is generally easy-to-use, although I might want more functionalities in the app. There are certain latencies in the app, which can be annoying. It would be more helpful if I can know if the device is not sending any traffic, or it is just simply late (e.g., adding a loading icon).} \textbf{(P2)}
    
    \textit{Make it possible to view past trends (a la net microscope) and scroll backwards in time, so I can get the context of how much traffic is regularly sent. Give me a global view of the worst offenders. Still some work to do on basic stability. It only works on devices that people have obviously ALREADY DECIDED TO BUY, which is a weird sample. Obviously, I don't have QR codes printed out on all of my household electronics.} \textbf{(P3)}
    
    \textit{I had difficulties trying to access the buttons, and the images seemed lagged a little. But the info was very useful overall.} \textbf{(P4)}
    
    \textit{Fix where the traffic and `learn more about the device' buttons once you've scanned the QR code. It's a bit awkward to have to hold the phone back up to the device. Maybe add the units (byte/kB) to the left hand side of the graph instead of above it for the traffic visualization.} \textbf{(P5)}
    
    \textit{The plots are not super-intuitive but I liked the representations in terms of text/pictures which is easier to comprehend. I would also be interested to see what advertisers the information is being leaked to. While the AR thing is cool, I would also like the option to just scroll through a list of devices. That ways I do not have to be close to the device and would also be able to monitor its activity when I am not close to the device. In fact, I would be interested in seeing the device communication (including interaction w/ advertisers) in that case.} \textbf{(P6)}

% that's all folks
\end{document}